%% file: main.tex
\documentclass{article}

\PassOptionsToPackage{x11names}{xcolor}
\usepackage[preprint]{colm2026_conference}
\usepackage{lineno}

\PassOptionsToPackage{numbers, compress, sort, square}{natbib}

\input{headers/math_commands.tex}

\input{headers/header-lqa}

\input{headers/header}

\input{headers/header-lqa-tail}

\pgfplotsset{compat=1.18}

\newcommand{\OurTitle}{Coding Agents Don't Know When to Act}
\title{\OurTitle}

\author{Thibaud Gloaguen\\
Department of Computer Science, ETH Zurich\\
\texttt{thibaud.gloaguen@inf.ethz.ch}
\And
Niels M{\"u}ndler\\
Department of Computer Science, ETH Zurich\\
\texttt{niels.mundler@inf.ethz.ch}
\AND
Mark M{\"u}ller\\
LogicStar.ai
\And
Veselin Raychev\\
LogicStar.ai
\And
Martin Vechev\\
Department of Computer Science, ETH Zurich
}

\begin{document}

\ifcolmsubmission
\linenumbers
\fi

\maketitle

\input{sections/abstract.tex}

\input{sections/introduction.tex}
\input{sections/methods.tex}
\input{sections/experiments.tex}
\input{sections/background.tex}

\input{sections/conclusion.tex}

\clearpage
\section*{Reproducibility Statement}
To ensure reproducibility of our results, we detail all our hyperparameters and additional configuration details in \cref{sec:experiments:setup,sec:additional_experiments}. We will release the code used to conduct the presented experiments. In the code, we pin requirements and evaluated LLM versions and coding agent harnesses. We report all used prompts in \cref{app:task-variant-prompts}. To ensure replicability, we focus on significant results and report all outcomes with a confidence interval at $95\%$.

\bibliography{references}
\bibliographystyle{colm2026_conference}

\input{sections/appendix}

\end{document}

%% file: headers/math_commands.tex
\usepackage{amsmath,amsfonts,bm}

\def\eqref#1{equation~\ref{#1}}

\def\1{\bm{1}}

\DeclareMathAlphabet{\mathsfit}{\encodingdefault}{\sfdefault}{m}{sl}
\SetMathAlphabet{\mathsfit}{bold}{\encodingdefault}{\sfdefault}{bx}{n}

%% file: headers/header-lqa.tex
\usepackage[utf8]{inputenc} %
\usepackage{lipsum} %

\usepackage[T1]{fontenc}

\usepackage{etoolbox}
\newbool{includeappendix}
\setbool{includeappendix}{true}

\input{headers/lqa/overfull}

\input{headers/lqa/comments}
\input{headers/lqa/abbreviations}

\input{headers/lqa/subfigures}

\input{headers/lqa/colors}

\input{headers/lqa/listing}

%% file: headers/lqa/overfull.tex
\ifdefined\isoverfull
\else
\fi

%% file: headers/lqa/abbreviations.tex
\newcommand{\eg}{e.g., }
\newcommand{\ie}{i.e., }

\newcommand{\swebench}{\textsc{SWE-bench Verified}\xspace}

\newcommand{\sorcar}{\textsc{SORCAR}\xspace}
\newcommand{\sonnet}{\textsc{Sonnet-4.6}\xspace}

\newcommand{\mini}{\textsc{GPT-5.4 mini}\xspace}
\newcommand{\gptfive}{\textsc{GPT-5.4}\xspace}
\newcommand{\gptfivecodex}{\textsc{GPT-5.3 Codex}\xspace}

\newcommand{\gptfivemini}{\textsc{GPT-5.4 mini}\xspace}

\newcommand{\gptfivenano}{\textsc{GPT-5 nano}\xspace}

\newcommand{\gemini}{\textsc{Gemini-3 Pro}\xspace}
\newcommand{\qwenbig}{\textsc{Qwen3.5-122B}\xspace}

\newcommand{\fixedbench}{\textsc{FixedBench}\xspace}

\newcommand{\novel}{\textsc{Novel}\xspace}
\newcommand{\tpartial}{\textsc{Partial}\xspace}
\newcommand{\resolved}{\textsc{Resolved}\xspace}

\newcommand{\bestcase}{\textsc{Best}\xspace}
\newcommand{\worstcase}{\textsc{Worst}\xspace}

\newcommand{\edit}{\textsc{Edit}\xspace}
\newcommand{\fix}{\textsc{Issue}\xspace}
\newcommand{\reproduceFix}{\textsc{Reproduce}\xspace}
\newcommand{\reproduceFixAbstain}{\textsc{Abstain or Fix}\xspace}

\newcommand{\ncategories}{10\xspace}

%% file: headers/lqa/subfigures.tex
\usepackage{subcaption}

%% file: headers/lqa/colors.tex
\definecolor{my-full-blue}{HTML}{1F77B4}

\definecolor{my-full-orange}{HTML}{FF7F0E}
\definecolor{my-extra-orange}{HTML}{7C4514}

\definecolor{my-full-green}{HTML}{2CA02C}

\definecolor{my-full-red}{HTML}{d62728}

\definecolor{my-full-purple}{HTML}{9467bd}

\definecolor{swebenchcolor}{HTML}{f8caca} %
\definecolor{planbenchcolor}{HTML}{d5f5db} %

\colorlet{my-blue}{my-full-blue!30}
\colorlet{my-orange}{my-full-orange!30}
\colorlet{my-green}{my-full-green!90}
\colorlet{my-red}{my-full-red!90}
\colorlet{my-purple}{my-full-purple!30}

\definecolor{mygreen}{HTML}{B5E3B5}
\colorlet{myred}{LightPink1}
\colorlet{myblue}{SteelBlue2}
\definecolor{mylightblue}{HTML}{B0DBFF}
\colorlet{myorange}{DarkOrange1}

\definecolor{darkblue}{RGB}{34,49,63}
\definecolor{lightblue}{RGB}{238,244,249}
\definecolor{accentblue}{RGB}{88,139,202}
\definecolor{lightgreen}{RGB}{220,245,230}
\definecolor{lightred}{RGB}{250,225,225}
\definecolor{darkgreen}{RGB}{40,167,69}
\definecolor{darkred}{RGB}{220,53,69}
\definecolor{textgray}{RGB}{120,120,120}
\definecolor{lightgray}{RGB}{240,240,240}   %
\definecolor{accentorange}{RGB}{127,17,144} %
\definecolor{gray2}{HTML}{FCFCFC}

%% file: headers/lqa/listing.tex
\usepackage{listings}

\usepackage{textcomp}

\usepackage{xcolor}

\usepackage[scaled=0.8]{beramono}

\definecolor{ckeyword}{HTML}{7F0055}
\definecolor{ccomment}{HTML}{3F7F5F}
\definecolor{cstring}{HTML}{2A0099}

\lstdefinestyle{numbers}{
	numbers=left,
	framexleftmargin=20pt,
	numberstyle=\tiny,
	firstnumber=auto,
	numbersep=1em,
	xleftmargin=2em
}

\lstdefinestyle{layout}{
	frame=none,
	captionpos=b,
}

\lstdefinestyle{comment-style}{
	morecomment=[l]//,
	morecomment=[s]{/*}{*/},
	commentstyle={\color{ccomment}\itshape},
}

\lstdefinestyle{string-style}{
	showstringspaces=false,%
}

\lstdefinestyle{keyword-style}{
	keywordstyle={\ttfamily\bfseries},
	morekeywords={
		function,
		constructor,
		int,
		bool,
		return,
		returns,
		uint
	},
	morekeywords = [2]{},
	keywordstyle = [2]{\text},
	sensitive=true,
}

\lstdefinestyle{input-encoding}{
	inputencoding=utf8,
	extendedchars=true,
	literate=
	{ℝ}{$\reals$}1%
	{→}{$\rightarrow$}1%
	{α}{$\alpha$}1%
	{β}{$\beta$}1%
	{λ}{$\lambda$}1%
	{θ}{$\theta$}1%
	{ϕ}{$\phi$}1%
}

\lstdefinestyle{escaping}{
	moredelim={**[is][\color{blue}]{\%}{\%}},
	moredelim={**[is][\color{pastelgreen}]{??}{??}},
	mathescape=true
}

\lstdefinestyle{default-style}{
	basicstyle=\fontencoding{T1}\ttfamily\footnotesize,
	style=numbers,
	style=layout,
	style=comment-style,
	style=string-style,
	style=keyword-style,
	style=input-encoding,
	style=escaping,
	tabsize=2,
	upquote=true
}

\lstdefinelanguage{BASIC}{
	language=C++,
	style=default-style
}[keywords,comments,strings]%

\lstdefinelanguage{JavaScript}{
morekeywords=[1]{break, continue, delete, else, for, function, if, in,
new, return, this, typeof, var, void, while, with, const, let},
morekeywords=[2]{false, null, true, boolean, number, undefined, string,
Array, Boolean, Date, Math, Number, String, Object},
morekeywords=[3]{eval, parseFloat, escape, unescape},
sensitive,
morecomment=[s]{/*}{*/},
morecomment=[l]//,
morecomment=[s]{/**}{*/}, %
morestring=[b]',
morestring=[b]"
}[keywords, comments, strings]
\definecolor{delim}{RGB}{20,105,176}
\definecolor{numb}{RGB}{106, 109, 32}
\definecolor{string}{rgb}{0.64,0.08,0.08}

\lstdefinelanguage{json}{
    showspaces=false,
    showtabs=false,
    breaklines=true,
    postbreak=\raisebox{0ex}[0ex][0ex]{\ensuremath{\color{gray}\hookrightarrow\space}},
    upquote=true,
    morestring=[b]",
    morecomment=[l]//,
    stringstyle=\color{string},
    literate=
     *{0}{{{\color{numb}0}}}{1}
      {1}{{{\color{numb}1}}}{1}
      {2}{{{\color{numb}2}}}{1}
      {3}{{{\color{numb}3}}}{1}
      {4}{{{\color{numb}4}}}{1}
      {5}{{{\color{numb}5}}}{1}
      {6}{{{\color{numb}6}}}{1}
      {7}{{{\color{numb}7}}}{1}
      {8}{{{\color{numb}8}}}{1}
      {9}{{{\color{numb}9}}}{1}
      {\{}{{{\color{delim}{\{}}}}{1}
      {\}}{{{\color{delim}{\}}}}}{1}
      {[}{{{\color{delim}{[}}}}{1}
      {]}{{{\color{delim}{]}}}}{1},
}
\definecolor{dkgreen}{rgb}{0,0.6,0}
\definecolor{dred}{rgb}{0.545,0,0}
\definecolor{dblue}{rgb}{0,0,0.545}
\definecolor{lgrey}{rgb}{0.9,0.9,0.9}
\definecolor{gray}{rgb}{0.4,0.4,0.4}
\definecolor{darkblue}{rgb}{0.0,0.0,0.6}
\lstdefinelanguage{cpp}{
      breaklines=true,               
      postbreak=\raisebox{0ex}[0ex][0ex]{\ensuremath{\color{gray}\hookrightarrow\space}},
      deletekeywords={...},          
      escapeinside={\%*}{*)},                  
      language=C++,                
      keywordstyle=\color{purple},  
      morekeywords={string,float}, 
      identifierstyle=\color{black},
      stringstyle=\color{blue},      
      showspaces=false,               
      showstringspaces=false,        
      showtabs=false,                
      tabsize=5,                     
    }

%% file: headers/header.tex
\usepackage[english]{babel}
\usepackage{xurl}
\usepackage[breaklinks=true]{hyperref}
\definecolor{darkpastelblue}{HTML}{0279AF}
\hypersetup{colorlinks,
      linkcolor=darkpastelblue,
      citecolor=darkpastelblue,
      urlcolor=darkpastelblue,
      filecolor=darkpastelblue}
\usepackage{algorithm}
\usepackage{algorithmicx}
\usepackage[noend]{algpseudocode}
\usepackage{color,soul}
\usepackage{ bbold }
\usepackage{multirow}
\usepackage{multicol}
\usepackage{caption}
\usepackage{tabularx,booktabs,xltabular}
\usepackage{graphicx}
\usepackage{tabu}
\usepackage{array}
\usepackage{siunitx}
\usepackage{subcaption}
\usepackage{fontawesome}
\usepackage{stackengine}
\usepackage{svg}
\usepackage{mathtools}
\usepackage{xspace}
\usepackage{wrapfig}
\usepackage{makecell}
\usepackage{placeins}
\usepackage{float}
\usepackage{pifont}
\usepackage{svg}
\usepackage[most]{tcolorbox}
\usepackage{array}
\usepackage{dsfont}
\usepackage{enumitem}
\usepackage[utf8]{inputenc}
\usepackage{textcomp}
\usepackage{amsthm}
\usepackage{amssymb,amsmath}
\usepackage[nounderscore]{syntax}
\usepackage{semantic}
\usepackage{rotating}
\newcolumntype{x}[2]{S[table-format=#1.#2,table-auto-round]}

\input{headers/lqa/tikz}
\usepackage{pgfplots}
\usepackage[capitalize,noabbrev]{cleveref}
\AtBeginDocument{
\crefname{appendix}{Appendix}{Appendices}
}

\usepackage{pifont}%

\definecolor{pastelgreen}{HTML}{059C05}
\definecolor{pastelred}{HTML}{FF7373}

\usepackage{setspace}

\newcommand{\ttt}[1]{\text{\texttt{#1}}}

\usepackage{scalerel}

\newcommand{\Hsquare}{%
  \text{\kern2\scriptspace\fboxsep=-.2pt\fbox{\rule{0pt}{1ex}\rule{1ex}{0pt}}\kern2\scriptspace}%
}

\algrenewcommand\alglinenumber[1]{\scriptsize #1\hspace{1mm}}
\newcounter{algoline}[algorithm]

\definecolor{PromptAccent}{HTML}{51a35f}
\definecolor{PromptVariantEdit}{HTML}{CCBB44}
\definecolor{PromptVariantFix}{HTML}{4477AA}
\definecolor{PromptVariantReproduce}{HTML}{66CCEE}
\definecolor{PromptVariantAbstain}{HTML}{228833}

\newcommand{\placeholder}[1]{%
  \tcbox[on line,
    colback=PromptAccent!9,
    colframe=PromptAccent!35,
    boxrule=0.35pt,
    arc=0.9mm,
    left=2.5pt,right=2.5pt,top=1pt,bottom=1pt
  ]{\ttfamily\footnotesize\{#1\}}%
}

\newtcolorbox{promptbox}[2][]{%
  enhanced,
  breakable,
  colback=white,
  colframe=#2!18,
  boxrule=0.55pt,
  arc=1.6mm,
  outer arc=1.6mm,
  left=10pt,right=10pt,
  top=8pt,bottom=8pt,
  borderline west={2.2pt}{0pt}{#2},
  drop shadow={black!12},
  fonttitle=\sffamily\bfseries\footnotesize,
  coltitle=white,
  colbacktitle=#2,
  title={#1},
  attach boxed title to top left={xshift=10pt,yshift*=-2mm},
  boxed title style={
    enhanced,
    arc=1.2mm,
    boxrule=0.35pt,
    colframe=#2,
    colback=#2,
    left=6pt,right=6pt,top=2pt,bottom=2pt
  },
  before skip=10pt,
  after skip=10pt
}

\newtcblisting{schemabox}{%
  enhanced,
  breakable,
  colback=black!2,
  colframe=black!14,
  boxrule=0.4pt,
  arc=1.2mm,
  left=8pt,right=8pt,
  top=6pt,bottom=6pt,
  listing only,
  listing options={
    basicstyle=\ttfamily\footnotesize,
    breaklines=true,
    columns=fullflexible,
    keepspaces=true,
    showstringspaces=false
  }
}

%% file: headers/lqa/tikz.tex
\usepackage{tikz}

\usetikzlibrary{arrows}
\usetikzlibrary{automata}
\usetikzlibrary{calc}
\usetikzlibrary{backgrounds}
\usetikzlibrary{decorations.markings}
\usetikzlibrary{decorations.pathmorphing}
\usetikzlibrary{decorations.pathreplacing}
\usetikzlibrary{fit}
\usetikzlibrary{patterns}
\usetikzlibrary{positioning}
\usetikzlibrary{shadows}
\usetikzlibrary{shapes}
\usetikzlibrary{shapes.geometric}
\usetikzlibrary{arrows.meta}
\usetikzlibrary{shadows.blur}

\definecolor{blue}{HTML}{347bc6}
\definecolor{green-underline}{HTML}{2de12c}
\definecolor{yellow-underline}{HTML}{ffd700}

\tikzstyle{block} = [
    minimum width=1cm,
    minimum height=0.5cm,
    align=center,
    fill=gray!30,
    rounded corners=5pt,
]

\tikzstyle{arrow} = [
	-{Latex[length=1.4mm, width=1.4mm]},
	draw=blue,
]

\tikzstyle{dashedline} = [
    draw=blue,
    dashed
]

\tikzstyle{line} = [
    draw=blue
]

\definecolor{lightblue}{RGB}{173,216,230} %

%% file: headers/header-lqa-tail.tex
\input{headers/lqa/references}

%% file: headers/lqa/references.tex
\usepackage[capitalize]{cleveref}

\crefformat{section}{\S#2#1#3}

\crefrangeformat{section}{\S#3#1#4\crefrangeconjunction\S#5#2#6}

\crefmultiformat{section}{\S#2#1#3}{\crefpairconjunction\S#2#1#3}{\crefmiddleconjunction\S#2#1#3}{\creflastconjunction\S#2#1#3}

\newcommand{\crefrangeconjunction}{--}

\crefname{listing}{Lst.}{listings}
\crefname{line}{Line}{Lines}
\crefname{appendix}{App.}{App.}

\newcommand{\app}[1]{%
	\ifbool{includeappendix}{\cref{#1}}{the appendix}%
}
\newcommand{\App}[1]{%
	\ifbool{includeappendix}{\cref{#1}}{The appendix}%
}

%% file: sections/abstract.tex
\begin{abstract}
Coding agents are increasingly deployed to autonomously maintain software, including to resolve user-reported issues: a bug report comes in and the agent creates a patch to address it. However, in any real-world deployment, they will encounter stale bug reports about issues that have already been resolved. Agents should recognize this and abstain from modifying the code to avoid accumulating technical debt.
To systematically evaluate whether current agents do so, we introduce \fixedbench{}, a code benchmark with 200 human-verified coding tasks in which no code changes are required, testing five recent models across four agent harnesses. We find that even state-of-the-art models fail, proposing undesirable changes (excluding tests and documentation) in $35$ to $65\%$ of cases. Explicit instructions to reproduce the issue before patching partially address this issue but introduce a new failure mode: when an issue is partially fixed, they abstain even though a patch would still be needed.
More broadly, our results indicate that LLMs fall prey to an action bias: they choose to act even if inaction would be appropriate. To break this pattern, inaction needs to be explicitly framed as a path to success, which highlights an overreliance on human guidance implicit in current training objectives.
\end{abstract}

%% file: sections/introduction.tex
\vspace{-2mm}
\section{Introduction}
\label{sec:introduction}
Coding agents are increasingly deployed for autonomous software maintenance.
In production settings, teams now routinely delegate issue resolution, code review, security scanning, and code editing to agents that operate with minimal supervision~\citep{ClaudeCodeDocsOverview, openaiCodex2026, googleGeminiCli, wang2025openhands}, and sometimes under full automation~\citep{CursorAutomations}.
A typical maintenance workflow involves an agent that receives a bug report, investigates the issue, writes a patch, and then submits it under full automation without any human review.
As this level of delegation transitions from individual tasks to continuous, event-driven pipelines, the quality of agent judgment becomes critical for long-term software maintenance.

A key aspect of this judgment involves knowing when \emph{not} to act.
In any real codebase, agents routinely encounter outdated bug reports, issues already fixed in a parallel pull request, or tickets that reference behavior that has since been patched. Such duplicate reports can make up to $49\%$ of all bug reports~\citep{duplication}.
A competent agent should recognize such situations, confirm that no code change is required, and move on without introducing technical debt.
This ability to abstain from undesirable changes is a prerequisite for autonomous maintenance, yet it has never been evaluated systematically.

\paragraph{Assessing Model Abstention Rates.} This work directly measures whether current coding agents can recognize when the code is already correct and abstain from making any meaningful changes.
To this end, we introduce \fixedbench{}, a random subset of 200 SWE-Bench Verified instances~\citep{swebench} in which the changes resolving the issue are already applied.
The expected behavior is simple: submit an empty patch, editing at most tests and documentation.
We evaluate five recent models across four agent harnesses and conduct thorough ablations to investigate failure modes and remediation strategies.

\paragraph{Models Fail to Abstain.} Our results show that even frontier agents and models fall prey to an action bias, incorrectly applying undesirable changes in $35-65\%$ of instances.
Manual trace analysis reveals two key factors: (i) models often never attempt to reproduce the reported bug before patching it, and (ii) even after realizing the issue is already fixed, the agent often feels compelled to modify the code.

\paragraph{Prompting Bias.} System prompts have a strong effect on this failure mode. Specifically, instructing the agent to edit the codebase exacerbates the problem, for example, almost halving the abstention rate of \mini from $60.5\%$ to $36.5\%$. In contrast, instructing it to first reproduce the problem and to abstain if it is already resolved alleviates the issue, for example, increasing the abstention rate for \mini to $88.5\%$. Interestingly, prompting the model only to reproduce the problem, without explicitly framing abstention as a successful outcome, has no effect or even an adverse effect, for example, reducing the abstention rate for \mini to $47.5\%$.
However, this prompting strategy is brittle: when the issue is only partially or incorrectly resolved, the same prompt causes models to over-abstain and, as a result, fail to fix code that genuinely needs repair.

\paragraph{Action Bias.} More broadly, our findings point to a gap between how agents are trained and what autonomous maintenance may require.
Current models are trained to act, \eg, to produce patches, rather than to decide whether action is required.
This behavior likely stems from the reinforcement learning post-training procedure of models, which is dominated by tasks that require a change~\citep{badertdinov2025swerebenchautomatedpipelinetask, mohamadi2025honesty, cambronero2025abstain}.
However, we argue that if agents that process thousands of changes per week systematically fail to recognize when code is already correct, the result is a steady accumulation of undesirable modifications to the codebase and increased technical debt.

\paragraph{Contributions.} The core contributions of our paper are as follows:
\begin{itemize}[leftmargin=1.5em, itemsep=2pt, topsep=2pt]
    \item A framework for generating benchmarks evaluating whether coding agents recognize that code is already correct, applied to \swebench.
    \item A thorough evaluation of multiple models and agent harnesses showing that none reliably abstain from making undesirable changes across a range of settings.
    \item An analysis of prompt-based remediation, showing that a verify-then-abstain prompt substantially improves abstention rates but introduces a complementary failure mode of over-abstention in the presence of incorrect prior patches.
    \item A large-scale trace analysis, identifying that agents feel compelled to edit the code even after realizing the issue is already resolved, suggesting that training objectives should reward verification, not just patch generation.
\end{itemize}

%% file: sections/methods.tex
\section{Benchmark}
\label{sec:method}

In this section, we outline the structure of our benchmark \fixedbench{} and how we leverage it to assess the capabilities of LLMs and coding agents to determine when a given issue is already resolved versus when it needs further work.

\subsection{Notation and Definitions}
\label{sec:notation_success}
We first introduce the notation to describe codebases, their test suites, and changes to these codebases in the form of patches.
A codebase $R$ is a set of files $f$. A patch $X$ is a set of tuples $(f, c)$ of a file and a changed, inserted, or removed line of content. $|X|$ counts the number of lines touched by a patch.
A patch can be applied to a codebase.
We denote a codebase $R$ after applying patch $X$ as $R \circ X$.
Several patches can be applied sequentially, i.e. $R \circ X \circ Y$ is the codebase $R$ after applying a first patch $X$ and then a second one $Y$. A set of tests $T$ can be used to check the correctness of a given codebase.

\newcommand{\myexec}{\mathrm{exec}}
A single test $s$ can either pass (P) or fail (F) after we execute it within the context of codebase $R$. If an error is thrown during a test's execution (either due to unexpected failure or explicit failed assertions), we consider it to have failed. Conversely, if the test executes without raising an error, we consider it to have passed. We define this process as an execution function: $\myexec(s, R) \in \{P, F\}$.

\subsection{Benchmark Overview}

Our evaluation is based on human-validated, real-world GitHub issues, in the style of SWE-Bench \citep{swebench}. For ease of overview, we briefly outline the construction process of \swebench{} \citep{openai2024swebenchverified}. The construction consists of four key steps.
\begin{enumerate}
    \item Scrape pull requests (PRs) from open-source Python repositories on GitHub.
    \item Filter PRs to only include those that were merged, resolved a GitHub issue, and made changes to at least one test file.
    \item Filter PRs to feature at least one test that fails in the pre-PR state $R$ and passes in the merged state $R \circ X^{*}$, removing PRs that result in installation or runtime errors.
    \item Filter instances with underspecified issues and overly restrictive test cases using manual annotation by human software engineers.
\end{enumerate}
This results in task instances consisting of a GitHub issue $I$, a pre-patch repository state $R$, a golden patch $X^{*}$ fixing the issue, and a set of golden reference tests $T^{*}$.

\subsection{Tasks}
We consider three tasks that each represent a stage in software development: \textsc{Novel}: resolve a novel issue, \textsc{Partial}: resolve an issue that was already partially or incorrectly addressed, and \textsc{Resolved}: process a stale issue that was already resolved.

Concretely, we initialize the coding agent in the following repository states for each task, given the pre-patch repository state $R$, a partial patch $X^{P}$, and the golden patch $X^{*}$.
\begin{itemize}
    \item $R_{\textsc{Novel}} = R$: The standard pre-patch repository state is used. This corresponds to the setup of standard SWE-Bench-like benchmarks. The golden test suite fails.
    \item $R_{\textsc{Partial}} = R \circ X^{P}$: An incorrect or insufficient patch generated by a weaker agent is applied to the pre-patch codebase. The golden test suite fails.
    \item $R_{\textsc{Resolved}} = R \circ X^{*}$: In the \textsc{Resolved} setting, the golden patch is applied to the pre-patch codebase. The golden test suite passes.
\end{itemize}

\begin{figure}
    \centering
    \includegraphics[width=\linewidth]{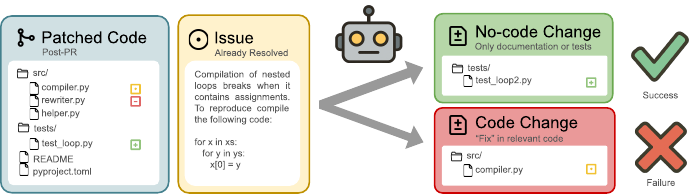}
    \caption{Overview of the main \fixedbench task (\resolved): The agent is tasked to resolve the user issue after it has already been resolved. Any (unnecessary) change to executable code is counted as failure.}
    \label{fig:overview-fixedbench}
\end{figure}

\subsection{Scenarios}

For all tasks, we consider two settings: a best-case scenario (\bestcase), and a worst-case scenario (\worstcase):

\paragraph{\bestcase} In the best-case scenario, the Git history is available with the last commit resolving the issue, and the agent operates in a fully set up execution environment (e.g., environment variables, packages).
There is no friction in running the repository test suite, and there are clear indications whether the bug is already fixed.
The \bestcase{} scenario represents a realistic setting where the agent is part of a deployment pipeline, for example, being automatically provided with submitted GitHub issues.
In \cref{sec:extended_evaluation} we also consider a best-case scenario in which the repository already contains a correct test file for the fixed issue.

\paragraph{\worstcase}
In the \worstcase{} scenario, the \texttt{.git} folder is removed and the environment is not set up, i.e., the agent needs to install the required packages on its own if it wants to run tests.
It allows us to assess the agent's performance under adverse conditions.

\subsection{Metrics}
\label{sec:metrics}

In order to successfully maintain code over long horizons, coding agents should produce patches $\hat{X}$ that are both minimal and correct. We introduce two metrics to assess this:

\paragraph{Resolution Rate} measures whether the produced patch resolves the issue. Concretely, we apply the patch $\hat{X}$ to the base repository of the task $R_{X}$ and execute the golden reference tests $T^{*}$. The task resolution metric \textsc{Pass} is 1 exactly when all golden reference tests pass, i.e., $\textsc{Pass} = \myexec(T^{*}, R_{X} \circ \hat{X}) = P$. The \emph{Resolution Rate} is its average across instances.

\paragraph{Abstention} measures whether a submitted patch makes no meaningful changes to the codebase.
We ignore changes that modify comments, documentation, or test cases and provide more detail on our filtering in \cref{sec:additional_method}. The metric \textsc{Abstention} is 1 exactly when the predicted patch (net comments and tests) is empty, i.e., $|\hat{X}| = 0$. We then compute the \emph{Abstention Rate} as its average. Depending on the task, we clarify the desired behavior by denoting it the \emph{Correct} (\resolved) or \emph{Incorrect} (\tpartial, \novel) abstention rate.

%% file: sections/experiments.tex
\vspace{-1mm}
\section{Experimental Evaluation}
\label{sec:experiments}
\vspace{-1mm}

\subsection{Setup}
\label{sec:experiments:setup}

\paragraph{Datasets}
While our benchmark generation approach can be applied to arbitrary SWE-Bench-like tasks, we construct \fixedbench from a random subset of 200 \swebench \citep{openai2024swebenchverified} instances.
Unless otherwise mentioned, we evaluate the \resolved{} setting, where we automatically commit the golden patch from the PR such that the issue described in the problem statement is resolved.

\paragraph{Metrics}
We measure the \emph{Resolution Rate}, the percentage of correctly resolved issues, and the \emph{Abstention Rate}, the percentage of patches without undesirable edits, for all instances, as described in \cref{sec:metrics}.
All reported error bars show $95$\% confidence intervals.
For statistical tests, we compute p-values with McNemar's test on the Abstention Rate.

\paragraph{Agent Instructions}
We prompt the agent with four different instructions:
\begin{enumerate}
  \item \fix: The instance description without additional information. This serves as the default baseline prompt.
  \item \edit: A prompt tasking the agent to edit the codebase to address the issue. This is an adverse condition reinforcing the action bias.
  \item \reproduceFix: First, reproduce the issue and then address it.
  \item \reproduceFixAbstain: First, reproduce the issue, and then either address the issue or abstain if there is nothing to fix.
\end{enumerate}
For the exact prompt templates, see \cref{app:task-variant-prompts}.

\begin{wrapfigure}[12]{r}{0.42\textwidth}
  \centering
  \vspace{-7mm}
  \includegraphics[width=0.42\textwidth]{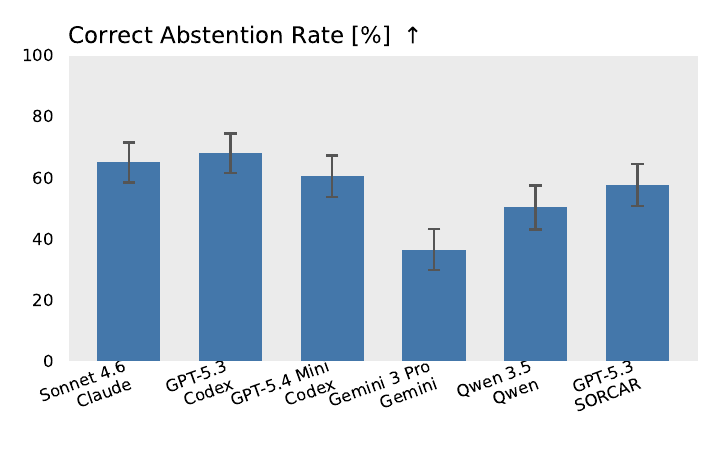}
  \vspace{-8mm}
  \caption{Correct abstention rate in the \bestcase{} scenario for the \fix prompt on the \resolved task.}
  \label{fig:best_case_by_model}
\end{wrapfigure}

\paragraph{Models}
For our main results, we use 4 closed models and 1 open-source model, \sonnet{}, \gptfivecodex{} (with xhigh thinking), \gptfivemini (with xhigh thinking), \gemini{}, and \qwenbig{}, with their corresponding agentic harnesses (\ie Claude-code for \sonnet{}, Codex for \gptfivemini and \gptfivecodex{}, Gemini-CLI for \gemini, and Qwen-Code for \qwenbig{}).
We also use the \sorcar harness with \gptfivecodex{} to evaluate open-source agentic frameworks.
For our ablations (\eg system prompt, scenario, and \tpartial), we use \sonnet{} and \gptfivemini{} only for cost reasons.
We defer details to \cref{sec:additional_experiments}.

\subsection{Main Results}

\input{table/main_table.tex}

\begin{wrapfigure}[17]{r}{0.45\textwidth}
  \vspace{-7mm}
  \centering
  \includegraphics[width=0.45\textwidth]{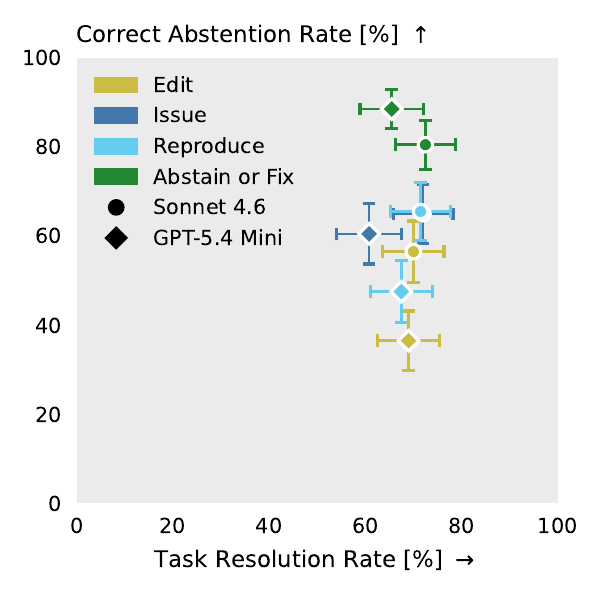}
  \vspace{-6mm}
  \caption{Correlation of correct-abstention rate and SWE-Bench task resolution across models and prompts.}
  \label{fig:best_case_by_variant_scatter}
\end{wrapfigure}

\paragraph{Agents Fall Victim to Action Bias}
Even under favorable conditions, i.e., with access to the full git history, a working execution environment, and the \fix prompt, coding agents struggle to recognize when code is already correct.
Across models and harnesses, agents apply unnecessary edits to already-fixed code in $35$ to $65\%$ of cases (see \cref{fig:best_case_by_model} and \cref{tab:fixed_code_scenario_results}).
Even the best-performing models \sonnet, \gptfivemini and \gptfivecodex do not achieve correct abstention rates above $67\%$, meaning that at least one in three stale issue reports results in spurious edits to working code.

\paragraph{Better Instructions Help}
We consider four different prompts to assess the impact of the agent's instructions on this failure mode and find a strong effect, visualized in \cref{fig:best_case_by_variant_scatter} and detailed in \cref{tab:fixed_code_scenario_results}.

Explicitly instructing the agent to edit the codebase to fix the issue (\edit) significantly exacerbates the action bias, reducing the (correct) abstention rate to $56.5\%$ and $36.5\%$ for \sonnet and \mini, respectively, down from $65.0\%$ and $60.5\%$.
In contrast, prompting the agents to first verify the issue is still present and abstain if it is not (\reproduceFixAbstain) improves correct-abstention rates significantly to $80.5\%$ and $88.5\%$, an uplift of $15.5$ ($p = 9.5 \times 10^{-3}$) and $28.0$ ($p = 2.3 \times 10^{-8}$) percentage points over their respective \fix baselines.
The \sorcar run with \gptfivecodex{} shows a similar improvement, from $57.6\%^{\pm6.9\%}$ under \fix{} to $83.5\%^{\pm5.1\%}$ under \reproduceFixAbstain{}.

Interestingly, prompting for reproduction \emph{without} the explicit option to abstain is insufficient.
The \reproduceFix{} variant yields only a negligible change for \sonnet ($65.0\%\to65.5\%$, $p = 1.0$) and even a notable degradation for \mini ($60.5\%\to47.5\%$, $p=6.5 \times 10^{-5}$), suggesting that reproduction steps alone do not help agents decide to hold back. Framing abstention as a successful outcome is what drives the improvement.

Importantly, these prompts do not compromise the agents' ability to resolve genuine bugs.
Across all four variants, SWE-Bench Verified resolution rates for both models remain within each other's confidence intervals, even improving performance in some cases~(\cref{fig:best_case_by_variant_scatter}).

\begin{wrapfigure}[17]{r}{0.45\textwidth}
  \vspace{-6mm}
  \centering
  \includegraphics[width=0.45\textwidth]{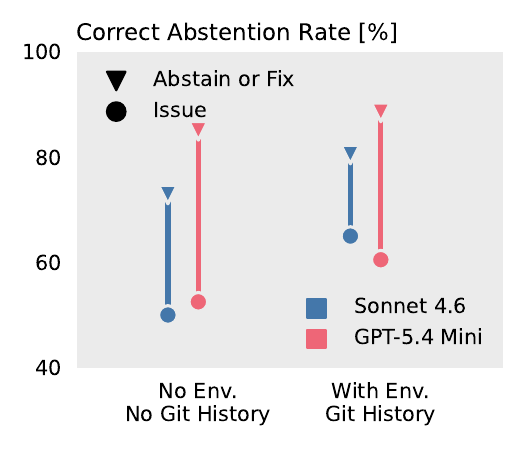}
  \vspace{-8mm}
  \caption{Correct abstention rate under \fix and \reproduceFixAbstain prompts across the \bestcase{} and \worstcase{} scenarios.}
  \label{fig:scenario_comparison}
\end{wrapfigure}

\paragraph{Adverse conditions.}
So far we have considered a best-case scenario, where agents had access to the git history, with the most recent commit addressing the issue at hand, and an already set up environment.
We now evaluate the effect of removing both the git history and the environment, visualizing results in \cref{fig:scenario_comparison} with details in \cref{tab:fixed_code_scenario_results}.
We observe that without an easy way to confirm task resolution, the correct abstention rate decreases from $65.0\%$ to $50.0\%$ and from $60.5\%$ to $52.5\%$ for \sonnet and \mini, respectively.
However, the \reproduceFixAbstain{} prompt remains effective, improving the non-edit rate from $50.0\%$ to $72.9\%$ ($p=1.1\times 10^{-8}$) and from $52.5\%$ to $85.0\%$ ($p=5.5\times 10^{-25}$) for \sonnet and \mini, respectively.
This almost matches the abstention rate in the \bestcase{} scenario, indicating that agents are capable of performing the necessary verification even under unfavorable conditions, provided they are explicitly instructed to do so.

\begin{wrapfigure}[16]{r}{0.45\textwidth}
  \vspace{-8mm}
  \centering
  \includegraphics[width=0.45\textwidth,trim=0 4mm 0 0,clip]{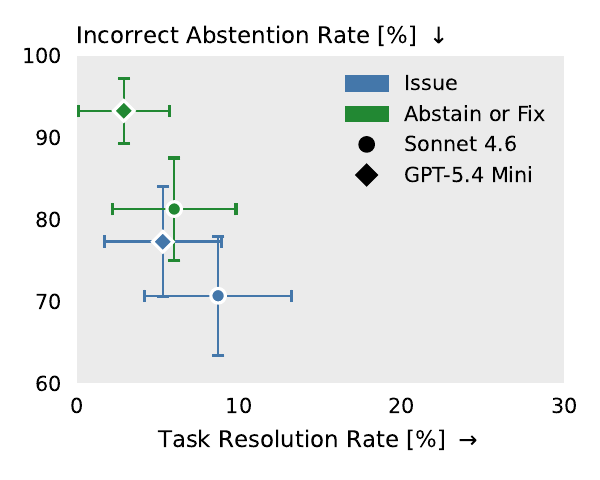}
  \captionsetup{skip=0.3pt}
  \caption{Incorrect abstention rate (higher is worse) vs.\ Resolution Rate on the \tpartial{} task, where a partial fix is applied and the agent should fix rather than abstain.}
  \label{fig:nano_swe_bench_scatter}
\end{wrapfigure}

\paragraph{The Problem with Instruction Changes.}

While our \reproduceFixAbstain instruction reduces unnecessary edits on already-fixed code without reducing the success rate on unmodified code, it introduces a symmetric failure mode: over-abstention on incorrect code that genuinely requires fixing.
We measure this with the \tpartial task on 150 instances where an incorrect patch (generated with \gptfivenano) is applied and the agent \emph{should} modify the codebase (see \cref{fig:nano_swe_bench_scatter} with details in \cref{tab:fixed_code_scenario_results}).
We report the \emph{incorrect abstention rate}, the fraction of instances where the agent incorrectly abstains from making changes, as the primary metric; higher values indicate worse performance.

Under the baseline \fix{} prompt, agents already exhibit a troubling degree of incorrect abstention: \sonnet and \mini only make any meaningful edit in $29.3\%^{\pm7.3\%}$ and $23.7\%^{\pm 6.7\%}$ of cases, resolving the issue only in $8.7\%^{\pm 4.5\%}$ and $5.3\%^{\pm 3.6\%}$ of cases, respectively.
This shows a concerning lack of ability to differentiate working from broken code, as resolution rates on the unmodified versions of the same subset of instances are much higher at $27.3\%^{\pm 7.9\%}$ and $24.4\%^{\pm 6.4\%}$, respectively.
Critically, the \reproduceFixAbstain{} prompt, which most effectively suppresses unnecessary edits on fixed code, \emph{increases} this incorrect abstention rate even further such that edits are only made in $18.7\%^{\pm6.2\%}$ and $6.4\%^{\pm4.3\%}$ of cases for \sonnet and \mini.

Prompt engineering thus trades one failure mode for another: it can suppress action bias on fixed code at the cost of inducing passivity on broken code.

\begin{wrapfigure}[15]{r}{0.4\textwidth}
  \vspace{-6mm}
  \centering
  \includegraphics[width=1.0\linewidth]{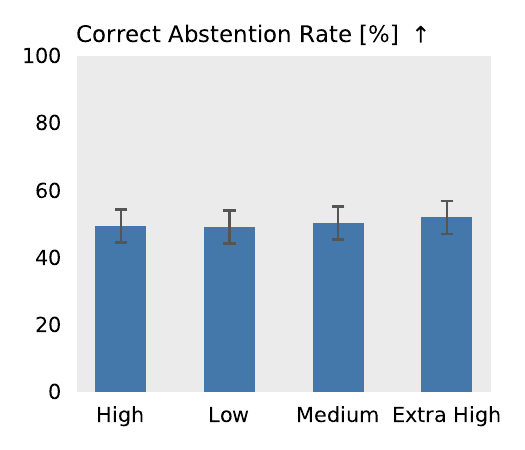}
  \vspace{-7mm}
  \caption{Abstention rate of \mini across thinking efforts.}
  \label{fig:thinking_by_strength}
\end{wrapfigure}

\paragraph{The (Lacking) Effect of Reasoning Effort.}
To investigate whether more reasoning helps agents avoid unnecessary edits, we vary the thinking effort of \mini across four levels under the \fix prompt, finding no significant effect (see \cref{fig:thinking_by_strength}).
While correct-abstention rates increase from $61.5\%$ to $63.3\%$, $64.3\%$, and $65.8\%$ for Low, Medium, High, and Extra High effort, respectively, this change (Spearman rank correlation of $r=1.0$ at $p=0.083$) is notably smaller than the confidence intervals of $\sim \pm7.0\%$ and the uplift resulting from a change of prompt.
This suggests that action bias is not a failure of deliberation that more compute can overcome, but rather a failure of task framing: agents need to be told that abstaining is a valid outcome and not just told to think harder.
This also suggests that they have not been trained to discern cases where the instructions (here fixing a non-existent bug) are wrong.

\subsection{Case Studies}

In this subsection, we analyze the traces of \sonnet across various settings.

\begin{wrapfigure}[20]{r}{0.6\textwidth}
  \vspace{-6mm}
  \centering
  \includegraphics[width=1.03\linewidth]{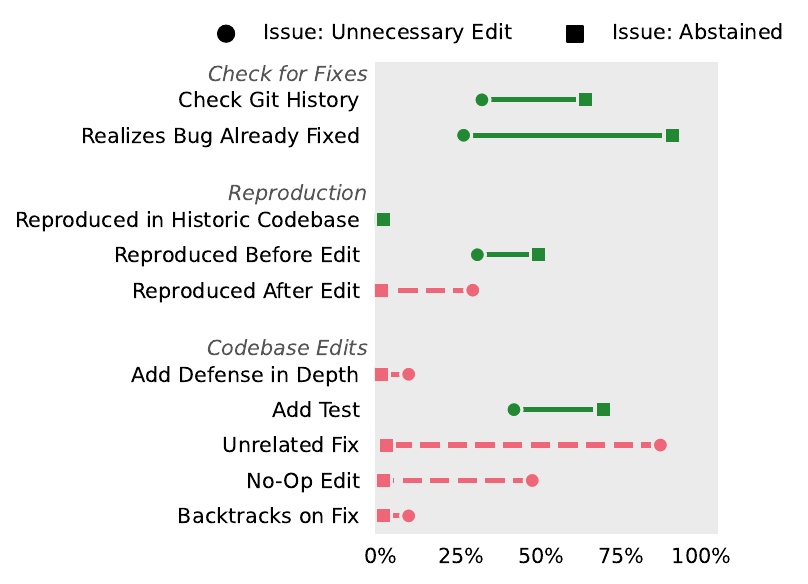}
  \vspace{-7mm}
  \caption{Comparison between traces leading to abstention vs.\ unnecessary edits in the best-case scenario. Green solid lines indicate higher frequency in the successful group; red dashed lines the reverse.}
  \label{fig:llm_judge_postpatch_diff_fix}
\end{wrapfigure}

\paragraph{Experimental Design}
We manually inspected 50 traces to define \ncategories (non-exclusive) behavioral categories, each corresponding to a relevant action of the agent (\eg looking at the git history or implementing a fix). See \cref{app:behavioral-categories} for detailed descriptions of all categories.
We then use \gptfive{} as a judge to identify (potentially multiple) behaviors in each trace. These traces consist of the provided tool calls and outputs, but not the thinking traces, as we found these to mislead the judge in a manual inspection.
We find high agreement between the human and LLM judge at $80.5\%$ precision and $75.0\%$ recall. For a detailed analysis, see \cref{sec:extended_evaluation}.

\paragraph{Why Agents (Don't) Abstain}
We examine what separates traces leading to correct abstention from those leading to unnecessary edits under the \fix prompt in the \bestcase scenario in \cref{fig:llm_judge_postpatch_diff_fix}.
We observe the two strongest predictors of correct abstention to be checking the git history ($63.8\%$ vs.\ $31.4\%$) and attempting to reproduce the issue before making any edits ($49.2\%$ vs.\ $30.0\%$): agents that pause to gather evidence are far more likely to recognize that the code is already correct (90.8 \% vs 25.7\%).

\begin{figure*}[t]
  \centering
  \begin{minipage}[t]{0.48\textwidth}
    \centering
    \hspace{-10mm}
    \includegraphics[width=1.10\linewidth]{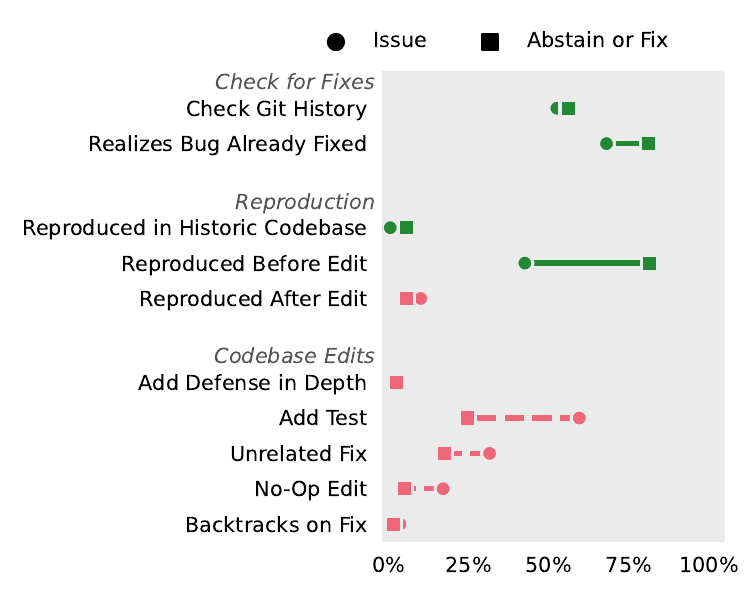}
    \caption{Comparison between the \fix and the \reproduceFixAbstain prompt. Green solid lines indicate behaviors more frequent in the \reproduceFixAbstain{} group; red dashed lines indicate the reverse.}
    \label{fig:llm_judge_by_variation}
  \end{minipage}
  \hfill
  \begin{minipage}[t]{0.48\textwidth}
    \centering
    \hspace{-10mm}
    \includegraphics[width=1.10\linewidth]{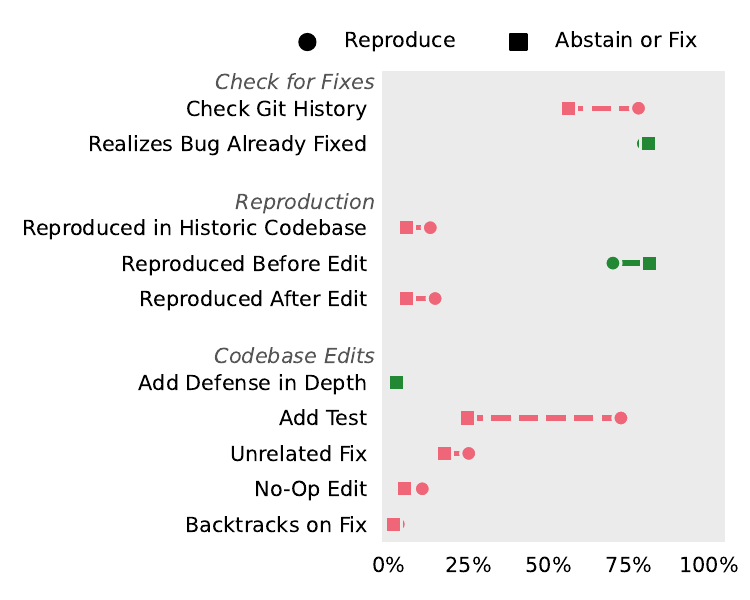}
    \caption{Comparison between the \reproduceFix and the \reproduceFixAbstain prompt. Green solid lines indicate behaviors more frequent in the \reproduceFixAbstain{} group; red dashed lines indicate the reverse.}
    \label{fig:llm_judge_by_repro_vs_abstain}
  \end{minipage}
\end{figure*}

Interestingly, agents that correctly abstained more frequently \emph{added a test} ($69.2\%$ vs.\ $41.4\%$), suggesting that they channel their action bias into verification rather than modification.
In contrast, unsuccessful traces are overwhelmingly characterized by agents modifying code unrelated to the issue they were supposed to resolve ($87.1\%$) or applying changes that do not meaningfully change behavior ($47.1\%$).

\paragraph{Why Prompting Helps}
Visualizing the behavioral differences between the \fix, \reproduceFix, and \reproduceFixAbstain prompts in \cref{fig:llm_judge_by_variation,fig:llm_judge_by_repro_vs_abstain}, we observe the most significant difference to be whether the agent attempts to reproduce the issue before making any edits ($81.5\%$ (\reproduceFixAbstain) and $70.0\%$ (\reproduceFix) vs $42.5\%$ \fix), leading the agent to realize the issue is already fixed far more frequently ($81.0\%$ and $79.5\%$ vs $68.0\%$).
However, realizing a bug is fixed alone does not prevent the agent from making undesirable edits, e.g., \reproduceFixAbstain and \reproduceFix realize this at almost the same rate of $81.0\%$ and $79.5\%$, respectively, while the correct abstention rate varies significantly ($80.5\%$ vs $65.5\%$).
Instead, the framing of abstention as successful task resolution drives a significant reduction in all forms of edits, both targeted fixes and broader refactors.
This suggests the key issue is not the agent's ability to reproduce an issue and, as a result, realize it is already resolved, but what it believes its success criteria are.

%% file: table/main_table.tex
\begin{table}[t]\centering
    \caption{The abstention rate (Abs.) and SWE-bench resolution rate (Res.) across tasks, scenarios, and prompts for selected models. The accuracy columns use the same raw SWE-bench accuracy for the corresponding prompt. All values are means $\pm$ 95\% binomial CIs.}
    \label{tab:fixed_code_scenario_results}

    \resizebox{0.85\linewidth}{!}{%
    \begingroup
    \setlength{\tabcolsep}{5pt}

    \begin{tabular}{lllcccc}
    \toprule
    \multirow{2}{*}{Task} & \multirow{2}{*}{Scenario} & \multirow{2}{*}{Prompt} & \multicolumn{2}{c}{\sonnet{}} & \multicolumn{2}{c}{\gptfivemini} \\
    \cmidrule(lr){4-5}\cmidrule(lr){6-7}
     &  &  & Abs. [\%] & Res. [\%] & Abs. [\%] & Res. [\%] \\
    \midrule
    \multirow{7}{*}{\resolved} & \multirow{4}{*}{\bestcase} & \fix{} & 65.0$^{\pm 6.6}$ & 72.0$^{\pm 6.2}$ & 60.5$^{\pm 6.8}$ & 60.8$^{\pm 6.8}$ \\
     &  & \edit{} & 56.5$^{\pm 6.9}$ & 70.0$^{\pm 6.4}$ & 36.5$^{\pm 6.7}$ & 69.0$^{\pm 6.4}$ \\
     &  & \reproduceFix{} & 65.5$^{\pm 6.6}$ & 71.5$^{\pm 6.3}$ & 47.5$^{\pm 6.9}$ & 67.5$^{\pm 6.5}$ \\
     &  & \reproduceFixAbstain{} & 80.5$^{\pm 5.5}$ & 72.5$^{\pm 6.2}$ & 88.5$^{\pm 4.4}$ & 65.5$^{\pm 6.6}$ \\
    \cmidrule(lr){2-7}
     & \multirow{3}{*}{\worstcase} & \fix{} & 50.0$^{\pm 7.0}$ & 72.0$^{\pm 6.2}$ & 52.5$^{\pm 6.9}$ & 60.8$^{\pm 6.8}$ \\
     &  & \edit{} & 42.5$^{\pm 7.3}$ & 70.0$^{\pm 6.4}$ & 30.5$^{\pm 6.4}$ & 69.0$^{\pm 6.4}$ \\
     &  & \reproduceFixAbstain{} & 72.9$^{\pm 6.2}$ & 72.5$^{\pm 6.2}$ & 85.0$^{\pm 4.9}$ & 65.5$^{\pm 6.6}$ \\
    \cmidrule(lr){1-7}
    \multirow{2}{*}{\tpartial} & \multirow{2}{*}{\bestcase} & \fix{} & 70.7$^{\pm 7.3}$ & 8.7$^{\pm 4.5}$ & 77.3$^{\pm 6.7}$ & 5.3$^{\pm 3.6}$ \\
     &  & \reproduceFixAbstain{} & 81.3$^{\pm 6.2}$ & 6.0$^{\pm 3.8}$ & 93.6$^{\pm 4.3}$ & 2.9$^{\pm 2.8}$ \\
    \bottomrule
    \end{tabular}
    \endgroup
    }
    \vspace{-0.1in}
\end{table}

%% file: sections/background.tex
\section{Related Work}
\label{sec:related-work}
\label{sec:background}

Our work connects several active research areas: software engineering benchmarks, LLM-based coding agents, and LLM topics like sycophancy and hallucinations.

\paragraph{Code generation benchmarks}
Early code generation benchmarks such as HumanEval~\citep{chen2021codex} and MBPP~\citep{MBPP2021} focus on isolated function synthesis.
\swebench~\citep{swebench} moved to software engineering tasks that edit an existing codebase, drawing tasks from real GitHub issues and pull requests.
More recent benchmarks~\citep{badertdinov2025swerebenchautomatedpipelinetask,vergopoulos2025automatedbenchmarkgenerationrepositorylevel,AGENTSmd} create tasks from more diverse, niche repositories where data contamination is less likely.
Our work directly builds on \swebench, but \emph{inverts} the standard evaluation setup: rather than presenting a codebase with an unresolved bug, we present one where a fix is already applied and measure whether agents recognize that the fix is correct and abstain.

Code generation benchmarks are fragile along multiple axes such as language and repository distribution, data contamination, and test adequacy.
\citet{riddell2024quantifyingcontaminationevaluatingcode} found substantial overlap between HumanEval/MBPP and open pretraining corpora.
These concerns motivated temporally-aware benchmarks: LiveCodeBench~\citep{jain2024livecodebench} collects post-training-cutoff problems, while SWE-rebench~\citep{badertdinov2025swerebenchautomatedpipelinetask} provides direct evidence of contamination, with a nearly 20 percentage point performance gap between \swebench{} and temporally-filtered tasks.
Our work still focuses only on Python, but completely sidesteps data contamination: even if a model has memorized the correct patch, the correct response in our setup is to produce \emph{no patch}, since the fix is already applied.

\paragraph{Automated program repair and test quality}
Bug-fixing software engineering tasks were previously known in the literature as automated program repair (APR)~\citep{zhang2023aprsurvey,jiang2026surveycodegen}, and were often defined as fixing buggy code given a failing test suite.
However, even outside of LLM-based approaches, a central concern in APR is \emph{patch overfitting}, i.e. generating patches that pass the tests but are semantically incorrect.
\citet{wang2021ase} found that most APR-generated patches overfit on some benchmarks, and \citet{mundler2024swt} showed that agent-generated fail-to-pass tests can substantially increase patch precision by serving as effective filters.
SWE-bench+~\citep{aleithan2024swebenchenhancedcodingbenchmark} revealed a related issue: many SWE-bench instances were incorrectly marked passing due to weak test cases.
Our benchmark requires LLMs to perform reasoning on whether a repair is correct and sufficient, a task strongly related to validating program repairs.

\paragraph{Abstention and sycophancy}
\citet{sharma2023understanding} demonstrated that RLHF-trained assistants systematically exhibit sycophantic behavior, preferring responses that match user beliefs over truthful ones.
Code agents exhibit what might be termed \emph{task-level sycophancy}: they comply with the implicit premise of the bug report rather than questioning whether that premise holds in the codebase.
The abstention literature provides frameworks~\citep{feng2024donthallucinateabstainidentifying} for deciding whether to act based on the knowledge ingested by the model.
As part of such a broad benchmark, \citet{kirichenko2025abstentionbenchreasoningllmsfail} found that strategic prompting can enhance abstention ability, but model reasoning does not improve abstention rates.
These broader benchmarks mirror our findings, but we quantify these effects in the context of code reasoning and bug-fixing tasks.

%% file: sections/conclusion.tex
\section{Limitations and Future Work}
We highlight three limitations of our work that present opportunities for future work:
First, our evaluation focuses on Python code from popular repositories. Future work may investigate repositories of different popularity \citep{AGENTSmd} and languages \citep{yang2025swesmith}.
Second, our benchmark does not capture the complexity of necessary patches.
Third, it remains open how to address the action bias failure mode without exacerbating the over-abstention failure mode in the presence of partial fixes.

\section{Conclusion}
\label{sec:conclusion}

We introduced \fixedbench{}, a benchmark measuring whether coding agents recognize when an issue is already resolved and does not require any edits.
Across recent models and agent harnesses, we find this capability to be lacking: even strong agents frequently modify already-correct code, revealing an action bias that is a critical failure mode in autonomous software maintenance, yet not captured by current coding benchmarks.
While we find that agent instructions that explicitly highlight the possibility to abstain can mitigate this failure mode, this comes at the cost of over-abstention on partially fixed code that still requires repair.

More generally, we argue that current agents are overreliant on the exact task framing and do not apply ``common sense'' by default, a gap that is critical for higher-autonomy applications.
We hope that \fixedbench{} raises awareness of this issue and helps shift evaluation and training toward this more realistic setting.

%% file: sections/appendix.tex
\clearpage
\appendix
\onecolumn

\section{Additional Method Details}
\label{sec:additional_method}

\begin{figure}
    \centering
    \includegraphics[width=\linewidth]{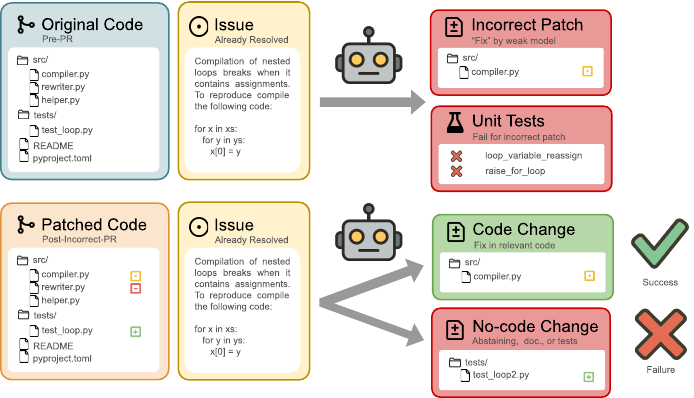}
    \caption{Overview of the \tpartial{} task and its construction: During construction, an agent is tasked with attempting to solve the issue, resulting in a patch that does not pass the golden test suite. During evaluation, an agent is tasked with resolving the user issue after the incorrect patch has been applied. If the model does not recognize the shortcoming of this patch, it counts as failure.}
    \label{fig:overview-nanobench}
\end{figure}

\paragraph{Constructing \tpartial{}}
We provide a visualization of the process to construct the \tpartial{} task for our benchmark in \cref{fig:overview-nanobench}.
Given the base repository state $R$ and the user issue $I$, we task a weak model (\gptfivenano) to produce a patch for this issue.
We then filter out instances in which the produced patch $\hat{X}$ passes the repository test suite.
In the \tpartial{} task setting, we then apply $\hat{X}$ to the repository state $R$ before initializing the agent.

\paragraph{Non-Meaningful Changes}
When assessing emptiness of patches, we explicitly filter out any changes to non-code and tests. This allows coding agents to modify function documentation or add and remove comments in code without being penalized, e.g., to remark that a certain issue has been resolved, or to add test cases validating the issue resolution.

Concretely, we preprocess each patch to filter out any patch lines where the modified file $f$ does not have a code suffix (e.g., \texttt{.py}, \ttt{.c}) or the path contains a test marker (e.g., \ttt{test\_}). Further, we remove all modified code lines $c$ starting with a comment marker (e.g., \ttt{\#, //}) or that we determine to be inside multi-line docstrings.

\section{Additional Experimental Details}
\label{sec:additional_experiments}

\paragraph{Models and Harness}
As explained in \cref{sec:experiments:setup}, we use 4 closed models, \sonnet{}, \gptfivecodex{} (with xhigh thinking), \gptfivemini (with xhigh thinking), \gemini{}, and 1 open-source model, \qwenbig{}, with their corresponding agentic harnesses (\ie Claude-code for \sonnet{}, Codex for \gptfivemini and \gptfivecodex{}, Gemini-CLI for \gemini{}, and Qwen-Code for \qwenbig{}).
In particular, we use the Anthropic API for \sonnet{} with Claude Code 2.1.81, the OpenAI API for \gptfivecodex{} and \gptfivemini{} with Codex 0.116.0, the Gemini API for \gemini{} with Gemini-CLI 0.34.0, and OpenRouter for \qwenbig{} with Qwen-Code 0.12.6.

\section{Extended Evaluation}
\label{sec:extended_evaluation}

\begin{wrapfigure}[17]{r}{0.43\textwidth}
    \centering
    \vspace{-4mm}
    \includegraphics[width=\linewidth]{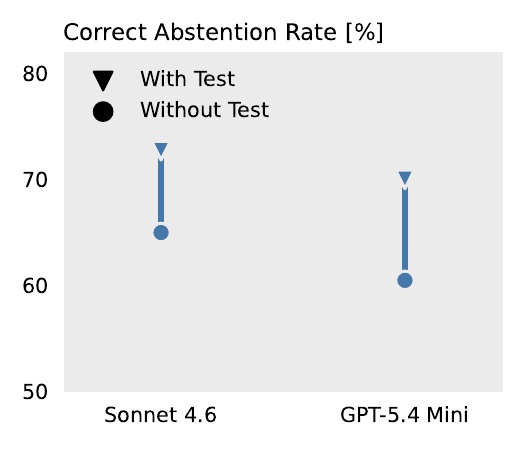}
    \vspace{-5mm}
    \caption{Effect of adding an already-existing correct test in the \bestcase{} scenario under the \fix{} prompt.}
    \label{fig:best_case_with_test_comparison}
\end{wrapfigure}

\paragraph{Effects of Already Existing Correct Test Files}
We additionally consider a \bestcase{} scenario in which the repository already contains a correct test file for the fixed issue.
This provides the agent with an even more direct verification signal: running the issue-specific test should indicate that the reported bug is already resolved.
As shown in \cref{fig:best_case_with_test_comparison}, this does improve performance under the \fix{} prompt, increasing the correct-abstention rate from $65.0\%$ to $72.7\%$ for \sonnet{} and from $60.5\%$ to $70.0\%$ for \gptfivemini{}.
However, the effect is not sufficient to solve the failure mode: even when a test for the already-fixed issue is present, both agents still fall substantially short of $100\%$ correct abstention and continue to make unnecessary edits in a large fraction of cases.

\begin{figure}[t]
  \centering
  \begin{minipage}{0.48\textwidth}
    \centering
    \includegraphics[width=\textwidth]{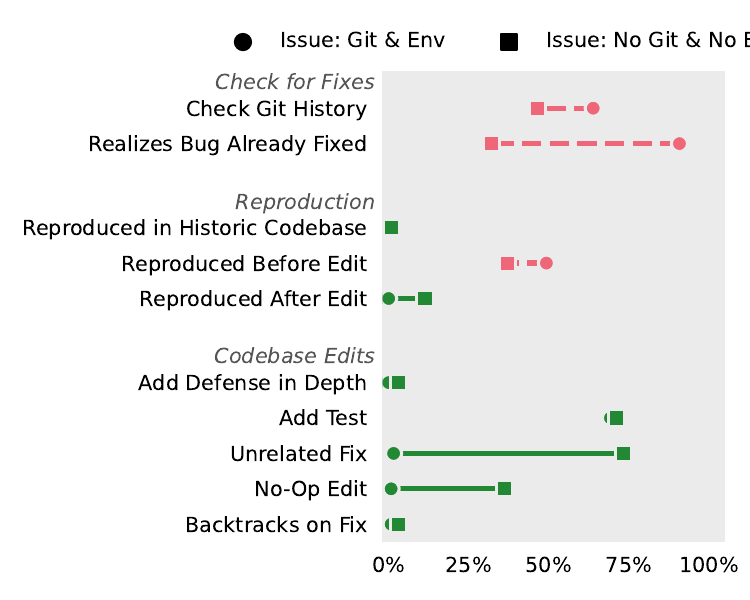}
    \caption{Behavioral comparison between the \bestcase{} and \worstcase{} scenarios under the \fix{} prompt, restricted to traces that correctly abstained from editing.}
    \label{fig:llm_judge_best_vs_worst_fix}
  \end{minipage}\hfill
  \begin{minipage}{0.48\textwidth}
    \centering
    \includegraphics[width=\textwidth]{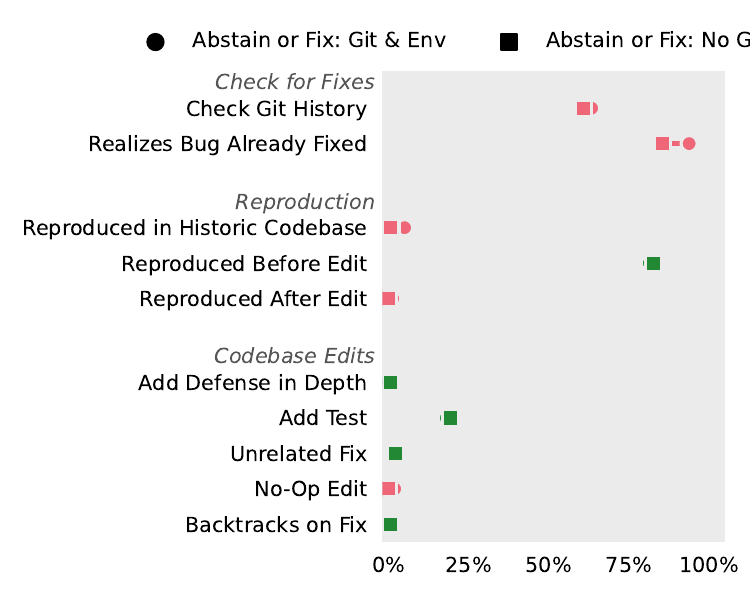}
\caption{Behavioral comparison between the \bestcase{} and \worstcase{} scenarios under the \reproduceFixAbstain{} prompt, restricted to traces that correctly abstained from editing.}
    \label{fig:llm_judge_best_vs_worst_abstain}
  \end{minipage}
\end{figure}

\paragraph{Behavioral Differences Across Scenarios}
Figures~\ref{fig:llm_judge_best_vs_worst_fix} and~\ref{fig:llm_judge_best_vs_worst_abstain} compare agent behavior between the \bestcase{} and \worstcase{} scenarios for the \fix{} and \reproduceFixAbstain{} prompts, respectively.
Under \reproduceFixAbstain{}, the scenario has little effect on agent behavior: among traces that correctly abstained, the rate of git history checks remains stable ($63.4\%$ vs.\ $60.7\%$), and the rate of reproducing the issue before editing barely changes ($81.4\%$ vs.\ $82.8\%$).
The \reproduceFixAbstain{} prompt therefore elicits the same verification strategy regardless of whether git history and the environment are set up.
In contrast, under the baseline \fix{} prompt, removing git history and not setting up the environment has a much more pronounced effect: agents check the git history and attempt to reproduce the issue before editing substantially less often.
This suggests that under the \fix{} prompt, agents rely heavily on the availability of explicit signals (git history, passing tests) to recognize that the code is already correct, and when those signals are stripped away, the verification behaviors that drive correct abstention largely disappear.

\begin{figure}[t]
    \centering
    \includegraphics[width=0.72\textwidth]{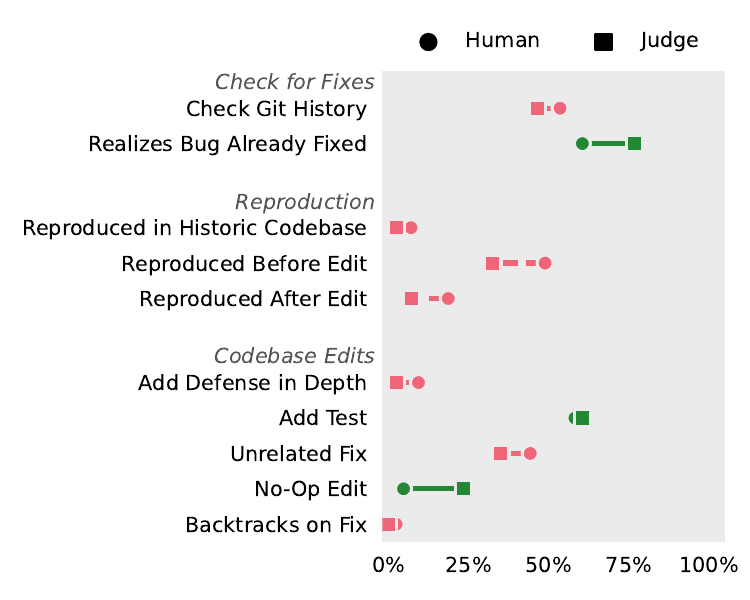}
    \caption{Comparison of per-category classification rates between manual annotations and the \gptfive{} judge on the 50 matched \sonnet{} traces in the \bestcase{} scenario under the \edit{} prompt.}
    \label{fig:llm_judge_vs_human_agreement}
\end{figure}

\paragraph{Evaluation of Trace Analysis Judge}
The trace analysis in \cref{sec:experiments} uses \gptfive{} as a multi-label judge over tool calls and tool outputs only, with the category definitions in \cref{app:behavioral-categories}.
To validate this setup, we compare the judge against manual annotations on 50 \sonnet{} traces from the \bestcase{} scenario under the \edit{} prompt and show the results in \cref{fig:llm_judge_vs_human_agreement}.
The overall patterns are similar, and pooling decisions across these categories gives a precision of $80.5\%$, a recall of $75.0\%$, and an F1 score of $77.6\%$ for the judge relative to the human annotations.
The largest disagreements come from the \emph{No-Op Edit} category (\ie, assessing whether the proposed fix is useless), which the judge assigns more often than humans do.
Such categories are indeed harder to classify, as they require some knowledge of the codebase to determine whether a given fix is useful.

\section{Behavioral Categories}
\label{app:behavioral-categories}

In this section, we list the category descriptions. We give these descriptions to the judge exactly.

\begin{enumerate}
  \item \textbf{Check Git History}: the agent looks at the git history to determine whether a fix already exists.
  \item \textbf{Realizes Bug Already Fixed}: the agent explicitly acknowledges that the reported bug is already completely fixed.
  \item \textbf{Reproduced in Historic Codebase}: the agent checks out or runs prior code to verify that the bug existed in an earlier version of the codebase.
  \item \textbf{Reproduced Before Edit}: the agent reproduces the issue before changing the main codebase file. Edits to tests or support code do not count against this category, and if main-code edits are reverted before reproduction, it still counts.
  \item \textbf{Reproduced After Edit}: the agent reproduces the issue only after changing the main codebase.
  \item \textbf{Add Defense in Depth}: the agent adds an additional fallback mechanism intended to catch the described problem.
  \item \textbf{Add Test}: the agent adds a test case for the issue.
  \item \textbf{Unrelated Fix}: the agent proposes and implements a fix to a similar issue rather than the one described in the problem.
  \item \textbf{No-Op Edit}: the agent implements a useless fix, \ie the functionality of the code before and after the edit is the same.
  \item \textbf{Backtracks on Fix}: the agent backtracks and cancels its changes.
\end{enumerate}

\section{Task-Variant Prompt Templates}
\label{app:task-variant-prompts}

For all task variants, we instantiate the prompt from a template where \placeholder{task} is replaced with the original task description.
The \fix{} variant uses the task description verbatim, while the other variants prepend or append additional instructions as shown below.

\begin{promptbox}[\fix{}]{PromptVariantFix}
\begin{schemabox}
{task}
\end{schemabox}
\end{promptbox}

\begin{promptbox}[\edit{}]{PromptVariantEdit}
\begin{schemabox}
Please edit the codebase to address the following task:

{task}
\end{schemabox}
\end{promptbox}

\begin{promptbox}[\reproduceFix{}]{PromptVariantReproduce}
\begin{schemabox}
{task}

Tackle the task like this:
First, reproduce the issue described above.
Then fix the issue and verify the reproduction now fails.
\end{schemabox}
\end{promptbox}

\begin{promptbox}[\reproduceFixAbstain{}]{PromptVariantAbstain}
\begin{schemabox}
{task}

Tackle the task like this:
First, reproduce the issue described above.
If the issue is not present (anymore), report that and don't make changes.
Otherwise fix the issue and verify the reproduction now fails.
\end{schemabox}
\end{promptbox}